\documentclass[letterpaper, 10 pt, conference]{ieeeconf}
\IEEEoverridecommandlockouts                              

\usepackage{amsmath,amsfonts,amssymb}%
\usepackage{bm}%
\usepackage{graphicx}%
\usepackage{cases}%
\usepackage{cite,url,citesort}
\usepackage{verbatim,latexsym}

\newtheorem{theorem}{Theorem}
\newtheorem{remark}{Remark}

\newtheorem{lemma}{Lemma}

\begin{document}
%
\title{\LARGE \bf
 Guaranteed Cost Dynamic Coherent Control for Uncertain Quantum Systems}

\author{Chengdi Xiang, Ian R. Petersen and Daoyi Dong
\thanks{Chengdi Xiang, Ian R. Petersen and Daoyi Dong are with the School of Engineering and Information Technology, University of New South Wales at the Australian Defence Force Academy, Canberra ACT 2600, Australia. {\tt\small \{elyssaxiang, i.r.petersen, daoyidong\}@gmail.com}}%
}


\maketitle

\begin{abstract}
This paper concerns a class of uncertain linear quantum systems  subject to quadratic perturbations in the system Hamiltonian. A small gain approach is used to evaluate the performance of the given quantum system. In order to get improved control performance, we propose two methods to design a coherent controller for the system. One is to formulate a static quantum controller by adding a controller Hamiltonian to the given system, and the other is to build a dynamic quantum controller which is directly coupled to the given system. Both controller design methods are given in terms of LMIs and a non-convex equality. Hence, a rank constrained LMI method is used as a numerical procedure. An illustrative example is given to demonstrate the proposed methods and also to make a performance comparison with different controller design methods. Results show that for the same uncertain quantum system, the dynamic quantum controller can offer an improvement in performance over the static quantum controller.
\end{abstract}

\IEEEpeerreviewmaketitle

\section{Introduction}
In recent years, there has been considerable interests focusing on quantum feedback control due to its applications in metrology, quantum optics, quantum computation, quantum communication and other quantum technologies [1]-[16]. Moreover, it has been recognized that quantum feedback control plays a vital role in manipulating a quantum mechanical system to achieve some pre-required closed loop properties such as stability \cite{Helon2006}, \cite{matt2008}, robustness \cite{Ian2012journal}, \cite{Valery}, entanglement \cite{entanglement} or other performance requirement \cite{Chengdi1}, \cite{Chengdi2}. In particular, linear quantum optics are widely studied in control areas and a quantum optical system can often be described by a set of linear quantum stochastic differential equations (QSDEs) \cite{matt2008}.

In the conventional picture of quantum feedback control, digital or analog electronic devices are often implemented as controllers \cite{Wiseman}. However, this tends to destroy quantum coherence and involves the destruction of quantum information in the process of making measurements. Hence, recent research has been focused on using a fully quantum system as a controller, which is referred to as a coherent controller, e.g., \cite{Chengdi1}, \cite{Nurdin2009}. Compared to measurement feedback control, the advantages of coherent control are to preserve quantum coherence, to achieve improved performance and to obtain a high speed processing bandwidth. However, when we consider coherent feedback control using the QSDE description, the issue of physical realizability of controllers arises \cite{matt2008}, \cite{Nurdin2009}. That is, in QSDEs framework, the state space matrices defining the coherent controller are required to satisfy certain conditions in order that the controller represents a physically meaningful quantum system and we call these kind of conditions physical realizability conditions. However, in this paper, we use an $(S,L,H)$ framework to define quantum systems, where $H$ is a Hamiltonian operator, $L$ is a vector of coupling operators and $S$ is a scattering matrix \cite{matt2009}, \cite{matt2010}. The quantity $L$ describes the interface between the system and the field, and the operator $H$ defines the self-energy of the system. Since the parameters $(S,L,H)$ already describe a physically realizable system, we do not need to be concerned about physical realizability conditions.

Quantum feedback controllers have been designed with a number of different techniques. For example, an $H^\infty$ synthesis approach \cite{matt2008} and a LQG method \cite{Nurdin2009} have been used to design a quantum controller for a class of linear quantum stochastic systems; \cite{Yanagisawa} used a transfer function method to analyze the robustness of feedback quantum systems. Nevertheless, few papers have considered quantum controller design based on an $(S,L,H)$ description. Based on the parameters $(S,L,H)$, we are going to design a guaranteed cost coherent controller not only to robustly stabilize the uncertain quantum system, but also to guarantee a specific level of performance for any admissible value of the uncertainties.

In the previous papers on quantum controller design \cite{matt2008}, \cite{Nurdin2009}, the coupling between the plant and the controller is via a field coupling which we call indirect coupling.
In the controller design parts of this paper, we use two different methods. One is to add controller Hamiltonian and the other is to construct a directly coupled quantum controller for the given system. Here, direct coupling refers to that two independent quantum systems may interact by exchanging energy \cite{wiseman1994}, \cite{guofeng} and this energy exchange is often described by an interaction Hamiltonian.

In this paper, the guaranteed cost coherent controller design is given in terms of LMI and nonlinear equality conditions. We use a nonlinear change of variables to convert the problem into a rank constrained LMI problem which can be solved using an alternating projections algorithm \cite{Orsi2}.

This paper is organized as follows.
In Section \ref{problem statement}, we define the nominal quantum system under consideration as a linear system using parameters $(S,L,H)$. Then an uncertain perturbation of the Hamiltonian is introduced in terms of a commutator decomposition and sector bound conditions in Section \ref{Perturbation of the Hamiltonian}.
In Section \ref{Performance Analysis}, the cost function for uncertain linear quantum systems subject to quadratic perturbation of the Hamiltonian is defined and a small gain type performance analysis result is presented. In Section \ref{static coherent controller}, we introduce a controller Hamiltonian and present a theorem to show the construction of this guaranteed cost quantum controller.
In Section \ref{Dynamic Coherent Controller}, a dynamic controller system is directly coupled to the uncertain quantum system. The corresponding controller design and numerical procedures are presented.
An illustrative example is presented to demonstrate the coherent controller methods in Section \ref{Illustrative Example}. We also make a performance comparison between the static coherent controller and the dynamic coherent controller.
Some conclusions are presented in Section \ref{conclusion}.
\section{System Description}\label{problem statement}
The open quantum system under consideration is an uncertain linear quantum system defined by parameters $(S,L,H)$, where $H$ refers to the system Hamiltonian and can be decomposed as $H=H_P+H_{un}$. Here, $H_P$ denotes a known nominal Hamiltonian and $H_{un}$ denotes a perturbation Hamiltonian contained in a specified set of Hamiltonians $\mathcal{W}$ \cite{Ian2012journal}.
 We assume that $H_P$ is in the form of
\begin{equation}\label{Hamiltonian}
H_P=\frac{1}{2}\left[\begin{array}{c c} q^T & p^T\end{array}\right]M_P\left[\begin{array}{l} q \\ p\end{array}\right]
\end{equation}
where $M_P$ is a real and symmetric Hermitian matrix with dimension $2n_P\times 2n_P$.
Here, $q$ is a vector of position operators and $p$ is a vector of momentum operators.
The commutation relations between position and momentum operators are described as follows
\begin{equation}\label{commutation_original}
\begin{split}
\left[\begin{array}{c}\left[\begin{array}{l}q\\p\end{array}\right] ,\left[\begin{array}{l}q\\p\end{array}\right]^T \end{array}\right]=&\left[\begin{array}{l}q\\p\end{array}\right] \left[\begin{array}{l}q\\p\end{array}\right]^T\\
&-\left(\begin{array}{c}\left[\begin{array}{l}q\\p\end{array}\right]\left[\begin{array}{l}q\\p\end{array}\right]^T \end{array}\right)^T\\
=&2i\theta=\Sigma,
\end{split}
\end{equation}
where $\theta=\left[\begin{array}{cr}0&I\\-I&0 \end{array} \right]$.

The coupling operator $L$ is of the form
\begin{equation}\label{single_coupling}
L=\left[\begin{array}{cc} N_1 & N_2\end{array}\right]\left[\begin{array}{l} q \\ p\end{array}\right],
\end{equation}
where $N_1\in\mathbb{C}^{m_P\times n_P}$ and $N_2\in\mathbb{C}^{m_P\times n_P}$.
We also have
\begin{equation}\label{coupling}
\left[\begin{array}{l} L \\ L^\#\end{array}\right]=N_P\left[\begin{array}{l} q \\ p\end{array}\right]
=\left[\begin{array}{cc} N_1 & N_2\\N_1^\# & N_2^\#\end{array}\right]\left[\begin{array}{l} q \\ p\end{array}\right].
\end{equation}

We consider self-adjoint ``Lyapunov" operators $V$ in the following form
\begin{equation}\label{v}
V=\left[\begin{array}{c c} q^T & p^T\end{array}\right]X\left[\begin{array}{l} q \\ p\end{array}\right]
\end{equation}
where $X\in \mathbb{R}^{2n_P\times2n_P}$  is a symmetric positive definite matrix.

Therefore, we define a set of non-negative self-adjoint operators $\mathcal{P}$ as follows
\begin{equation}
\mathcal{P}=\left\{\begin{array}{c} V\ \textnormal{ of the form} \ (\ref{v})\ \textnormal{where} \ X>0\ \textnormal {is a}\\ \textnormal{symmetric positive definite matrix}\end{array} \right\}.
\end{equation}

To proceed, we also define the corresponding generator operator as follows
\begin{equation}
\mathcal{G}(\bold{X})=-i[\bold{X},H]+\mathcal{L}(\bold{X}),
\end{equation}
where $\mathcal{L}(\bold{X})=\frac{1}{2}L^\dag[\bold{X},L]+\frac{1}{2}[L^\dag,\bold{X}]L$. Here, the notation $^\dag $ stands for the adjoint transpose of a vector of operators and $[\bold{X},H]=\bold{X}H-H\bold{X}$ describes the commutator between two operators. The following lemma will be used in the main results presented in this paper.

\begin{lemma}\label{1lemma}
\cite{Valery} Consider an open quantum system defined by $(S, L, H)$ and suppose there exist non-negative self-adjoint operators $V$ and $W$ on the underlying Hilbert space such that
\begin{equation}\label{lemma1}
\mathcal {G}(V)+W\leq \lambda
\end{equation}
where $\lambda$ is a real number. Then for any plant state, we have
\begin{equation}
\limsup \limits_{T\to \infty} \frac{1}{T}\int_0^T\langle W(t)\rangle dt\leq\lambda.
\end{equation}
Here $W(t)$ denotes the Heisenberg evolution of the operator $W$ and $\langle\cdot\rangle$ denotes quantum expectation; e.g., see \cite{Valery} and \cite{matt2010}.
\end{lemma}
\section{Perturbation of the Hamiltonian}\label{Perturbation of the Hamiltonian}
In this section, we introduce a perturbation for the quantum system under consideration. First, we define the perturbation of the Hamiltonian in terms of a commutator decomposition. Then, we introduce the formulation of the quadratic perturbation in the system Hamiltonian.
\subsection{Commutator Decomposition}
For the set of non-negative self-adjoint operators $\mathcal{P}$ and given real parameters $\gamma>0$, $\delta\geq 0$, a particular set of perturbation Hamiltonians $\mathcal{W}_1$ is defined in terms of the commutator decomposition
\begin {equation}\label{decomposition1}
[V,H_{un}]=[V, z^T]w-w^T[z,V]
\end{equation}
for $V\in \mathcal{P}$, where $w$ and $z$ are given real vectors of operators.
$\mathcal{W}_1$ is then defined in terms of sector bound condition:
\begin{equation}\label{sector1}
w^T w\leq\frac{1}{\gamma^2}z^T z+\delta.
\end{equation}

We define
\begin{equation}
\mathcal{W}_1=\left\{\begin{array}{c}H_{un} :\exists \ w, z\ \textnormal{such that} \ (\ref{decomposition1})\  \textnormal{and} \ (\ref{sector1})\ \\ \ \textnormal{are satisfied}\ \forall \ V\in \mathcal{P}\end{array} \right\}.
\end{equation}

\begin{lemma}\label{lemmaw1}
Consider an open quantum system $( S, L, H)$ where $H=H_P+H_{un}$ and $H_{un} \in \mathcal{W}_1$, and the set of non-negative self-adjoint operators $\mathcal{P}$. If there exists a $V \in \mathcal{P}$ and a real constant $\lambda\geq0$ such that
\begin{equation}\label{lemmaw1equation}
-i[V,H_P]+\mathcal{L}(V)+[V,z^T][z,V]+\frac{1}{\gamma^2}z^T z+W\leq\lambda,
\end{equation}
then
\begin{equation}
\limsup \limits_{T\rightarrow \infty} \frac{1}{T}\int_0^T\langle W(t)\rangle dt\leq\lambda+\delta, \forall t\geq0.
\end{equation}
\end{lemma}
\quad\\

\emph{Proof:}
As we know $V\in \mathcal{P}$ and $H_{un} \in \mathcal{W}_1$,
\begin{equation}\label{e9}
\mathcal{G}(V)=-i[V,H_P]+\mathcal{L}(V)-i[V, z^T]w+iw^T[z,V].
\end{equation}
Since $V$ is symmetric $[V,z^T]^\dag=[z,V]$.
Therefore,
\begin{equation}\label{e10}
\begin{split}
0&\leq([V,z^T]-iw^T)([V,z^T]-iw^T)^\dag\\
&=[V,z^T][z,V]+i[V,z^T]w-iw^T[z,V]+w^Tw.
\end{split}
\end{equation}
Substituting (\ref{e10}) into (\ref{e9}) and using the sector bound condition (\ref{sector1}), the following inequality is obtained:
\begin{equation}
\mathcal {G}(V)\leq-i[V,H_P]+\mathcal{L}(V)+[V,z^T][z,V]+\frac{1}{\gamma^2}z^T z+\delta.
\end{equation}
It follows from (\ref{lemmaw1equation}) that $\mathcal {G}(V)+W\leq \lambda+\delta.$
Consequently, the result follows from Lemma \ref{1lemma}.
\hfill $\Box$

\subsection{Quadratic Hamiltonian Perturbation}
A set of quadratic perturbation uncertainties is defined in the following form
\begin{equation}\label{h2form1}
H_{un}=\frac{1}{2}\left[\begin{array}{c c} \zeta_q^T & \zeta_p^T\end{array}\right]\Delta\left[\begin{array}{l} \zeta_q \\ \zeta_p\end{array}\right]
\end{equation}
where $\Delta^T=\Delta$. We also have the relationship
\begin{equation}\label{z}
z=\left[\begin{array}{l} \zeta_q \\ \zeta_p\end{array}\right]=E \left[\begin{array}{l} q \\ p\end{array}\right].
\end{equation}
Hence, we write
\begin{equation}\label{perturbation_equation}
H_{un}=\frac{1}{2}\left[\begin{array}{c c} q^T & p^T\end{array}\right]E^T\Delta E\left[\begin{array}{l} q \\ p\end{array}\right].
\end{equation}

The matrix $\Delta$ is subject to the norm bound
\begin{equation}\label{delta_bound}
\| \Delta \|\leq\frac{2}{\gamma}
\end{equation}
where $\|.\|$ refers to the matrix induced norm and we have the following definition

\begin{equation}
\mathcal{W}_2=\left\{\begin{array}{c}H_{un} \textnormal { of the form (\ref{h2form1}) such that }\\
\textnormal{condition (\ref{delta_bound}) is satisfied}\end{array} \right\}.
\end{equation}

\begin{lemma}\label{lemma_perturbation}
For any set of self-adjoint operators $\mathcal{P}$,
\begin{equation}
\mathcal{W}_2\subset\mathcal{W}_1
\end{equation}
\end{lemma}

\emph{Proof}: Given any $H_{un}\in\mathcal{W}_2$, let
\begin{equation}
w=\frac{1}{2}\left[\begin{array}{cc}\Delta_{11}&\Delta_{12}\\ \Delta^T_{12}&\Delta_{22}\end{array}\right]\left[\begin{array}{l} \zeta_q \\ \zeta_p\end{array}\right]
=\frac{1}{2}\Delta E\left[\begin{array}{l} q \\ p\end{array}\right]
\end{equation}
and
\begin{equation}
z=\left[\begin{array}{l} \zeta_q \\ \zeta_p\end{array}\right]=E\left[\begin{array}{l} q \\ p\end{array}\right].
\end{equation}
Hence,
\begin{equation}
H_{un}=w^Tz=\frac{1}{2}\left[\begin{array}{c c} q^T & p^T\end{array}\right]E^T\Delta E\left[\begin{array}{l} q \\ p\end{array}\right].
\end{equation}
Then, for any $V\in\mathcal{P}$,
\begin{equation}
\begin{split}
[V,z^T]w=&\frac{1}{2}\left(\begin{array}{l} V\zeta_q^T\Delta_{11}\zeta_q+V\zeta_q^T\Delta_{12}\zeta_p \\ +V\zeta_p^T\Delta_{12}^T\zeta_q+V\zeta_p^T\Delta_{22}\zeta_p\end{array}\right)\\
&-\frac{1}{2}\left(\begin{array}{l} \zeta_q^TV\Delta_{11}\zeta_q+\zeta_q^TV\Delta_{12}\zeta_p \\ +\zeta_p^TV\Delta_{12}^T\zeta_q+\zeta_p^TV\Delta_{22}\zeta_p\end{array}\right)
\end{split}
\end{equation}
Also,
\begin{equation}
\begin{split}
w^T[z,V]=&\frac{1}{2}\left(\begin{array}{l} \zeta_q^T\Delta_{11}\zeta_qV+\zeta_q^T\Delta_{12}\zeta_pV \\ +\zeta_p^T\Delta_{12}^T\zeta_qV+\zeta_p^T\Delta_{22}\zeta_pV\end{array}\right)\\
&-\frac{1}{2}\left(\begin{array}{l} \zeta_q^TV\Delta_{11}\zeta_q+\zeta_q^TV\Delta_{12}\zeta_p \\ +\zeta_p^TV\Delta_{12}^T\zeta_q+\zeta_p^TV\Delta_{22}\zeta_p\end{array}\right)
\end{split}
\end{equation}
Hence,
\begin{equation}
\begin{split}
&[V,z^T]w-w^T[z,V]\\
&=\frac{1}{2}\left(\begin{array}{l} V\zeta_q^T\Delta_{11}\zeta_q+V\zeta_q^T\Delta_{12}\zeta_p \\ +V\zeta_p^T\Delta_{12}^T\zeta_q+V\zeta_p^T\Delta_{22}\zeta_p\end{array}\right)\\
&-\frac{1}{2}\left(\begin{array}{l} \zeta_q^T\Delta_{11}\zeta_qV+\zeta_q^T\Delta_{12}\zeta_pV \\ +\zeta_p^T\Delta_{12}^T\zeta_qV+\zeta_p^T\Delta_{22}\zeta_pV\end{array}\right)\\
&=VH_{un}-H_{un}V=[V,H_{un}]
\end{split}
\end{equation}
Also,
\begin{equation}
\frac{1}{4}\left[\begin{array}{c c} \zeta_q^T & \zeta_p^T\end{array}\right]\Delta\Delta\left[\begin{array}{l} \zeta_q \\ \zeta_p\end{array}\right]\leq\frac{1}{\gamma^2}\left[\begin{array}{c c} \zeta_q^T & \zeta_p^T\end{array}\right]\left[\begin{array}{l} \zeta_q \\ \zeta_p\end{array}\right]
\end{equation}
Therefore, we have
$\mathcal{W}_2\subset\mathcal{W}_1.$
\hfill $\Box$

\section{Performance Analysis}\label{Performance Analysis}
In this section, we evaluate the performance of the given uncertain linear quantum system. First,
we need to define the associated cost function for a quantum system as
\begin{equation}
J=\limsup \limits_{T\to \infty} \frac{1}{T}\int_0^T\langle \left[\begin{array}{c c} q^T & p^T\end{array}\right]R\left[\begin{array}{l} q \\ p\end{array}\right]\rangle dt
\end{equation}
where $R>0$.
We denote that
\begin{equation}
W=\left[\begin{array}{c c} q^T & p^T\end{array}\right]R\left[\begin{array}{l} q \\ p\end{array}\right].
\end{equation}

In order to introduce our results on performance analysis, we require the following algebraic identities.

\begin{lemma}\label{algebra}
Consider $V\in\mathcal{P}, H_P$ is of the form (\ref{Hamiltonian}) and $L$ is of the form (\ref{coupling}). Then we have
\begin{equation}
[V, H_P]= \left[\begin{array}{l}q\\p\end{array}\right]^T(X\Sigma M_P-M_P\Sigma X) \left[\begin{array}{l}q\\p\end{array}\right],
\end{equation}
and
\begin{align}
\mathcal{L}(V)=&-\frac{1}{2} \left[\begin{array}{l}q\\p\end{array}\right]^T(N_P^\dag JN_P\Sigma X+X\Sigma N_P^\dag JN_P)\left[\begin{array}{l}q\\p\end{array}\right]\nonumber\\
&+\textnormal{Tr}(X\Sigma N_P^\dag \left[\begin{array}{cc} I&0\\0&0\end{array}\right]N_P\Sigma),
\end{align}
where $J=\left[\begin{array}{cr}I&0\\0&-I\end{array}\right]$.
Moreover,
\begin{equation}
[\left[\begin{array}{l}q\\p\end{array}\right],\left[\begin{array}{l}q\\p\end{array}\right]^T X\left[\begin{array}{l}q\\p\end{array}\right]]=2\Sigma X\left[\begin{array}{l}q\\p\end{array}\right].
\end{equation}
\end{lemma}
\emph{Proof:} The proof of these identities follows via straightforward but tedious calculations using (\ref{commutation_original}).
\hfill $\Box$
\begin{lemma}\label{zlemma}
For $V\in\mathcal{P}$ and $z$ defined in (\ref{z}),
\begin{equation}
[z,V]=2E\Sigma X\left[\begin{array}{l} q\\p \end{array}\right],
\end{equation}

\begin{equation}
[V,z^T][z,V]=4\left[\begin{array}{l} q\\p \end{array}\right]^T X\Sigma^\dag E^T E\Sigma X\left[\begin{array}{l} q\\p \end{array}\right],
\end{equation}

\begin{equation}
z^T z=\left[\begin{array}{l} q\\p \end{array}\right]^T E^T E\left[\begin{array}{l} q\\p \end{array}\right].
\end{equation}
\end{lemma}
\emph{Proof:} The result follows from Lemma \ref{algebra}. \hfill $\Box$

We now in a position to present our main result in this section.

\begin{theorem}\label{theorem1}
Consider an uncertain quantum system $(S, L, H)$, where $H=H_P+H_{un}$, $H_P$ is in the form of (\ref{Hamiltonian}),
$L$ is of the form (\ref{single_coupling}) and $H_{un}\in\mathcal{W}_2$. If $A=-i\Sigma M_P-\frac{1}{2}\Sigma N_P^\dag JN_P$ is Hurwitz, and
\begin{equation}\label{riccati}
\left[\begin{array}{cc}A^T X+XA+\frac{E^T E}{\gamma^2}+R&2X\Sigma^\dag E^T\\ 2E\Sigma X&-I\end{array}\right]<0
\end{equation}
has a symmetric solution $X>0$, then
\begin{align}
J&=\limsup \limits_{T\to \infty} \frac{1}{T}\int_0^T\langle W(t)\rangle dt\nonumber\\
&=\limsup \limits_{T\to \infty} \frac{1}{T}\int_0^T\langle \left[\begin{array}{c c} q^T & p^T\end{array}\right]R\left[\begin{array}{l} q \\ p\end{array}\right]\rangle dt\leq\lambda+\delta
\end{align}
where
\begin{equation}
\lambda=\text{Tr}(X\Sigma N_P^\dag \left[\begin{array}{cc} I&0\\0&0\end{array}\right]N_P\Sigma).
\end{equation}
\end{theorem}
\emph{Proof:} The proof is similar to that in Theorem 1 of \cite{Chengdi1}, and a detailed proof is omitted.
\hfill $\Box$

\section{Static Coherent Controller}\label{static coherent controller}
In this section, we aim to design a coherent guaranteed cost controller for the given uncertain quantum system by adding a controller Hamiltonian $H_{K1}$. As there are no additional dynamic variables defined by this Hamiltonian, we refer this kind of quantum controller as a static coherent controller.

The controller Hamiltonian $H_{K1}$ is assumed to be in the form
\begin{equation}\label{H3}
H_{K1}=\frac{1}{2}\left[\begin{array}{c c} q^T & p^T\end{array}\right]F^T KF\left[\begin{array}{l} q \\ p\end{array}\right]
\end{equation}
where $F\in\mathbb{R}^{2n_{P}\times2n_P}$ and $K\in\mathbb{R}^{2n_{P}\times2n_P}$ is a symmetric matrix.
An associated cost function $J$ is defined in the following form
\begin{equation}
J=\limsup \limits_{T\to \infty} \frac{1}{T}\int_0^\infty \langle\left[\begin{array}{l} q \\ p\end{array}\right]^T (R+ \rho F^T KF F^T KF)\left[\begin{array}{l} q \\ p\end{array}\right]\rangle dt,
\end{equation}
where $\rho\in(0,\infty)$ is a weighting factor. We have

\begin{equation}\label{nondynamic_wt}
W=\left[\begin{array}{l} q \\ p\end{array}\right]^T (R+\rho F^T KF F^T KF)\left[\begin{array}{l} q \\ p\end{array}\right].
\end{equation}
\subsection{Controller Design}

\begin{theorem}\label{nondynamic_theorem}
Consider an uncertain quantum system $(S, L, H)$, where $H=H_P+H_{un}+H_{K1}$, $H_P$ is in the form of (\ref{Hamiltonian}),
$L$ is of the form (\ref{single_coupling}) and $H_{un}\in\mathcal{W}_2$, and the controller Hamiltonian $H_{K1}$ is in the form of (\ref{H3}). If there exists symmetric matrices $K$, $X>0$ and $Y=K F\theta^T X$ such that
\begin{equation}\label{nondynamic_inequality}
\left[\begin{array}{ccc}B&4X\theta^TE^T&F^T KF\\4E\theta X&-I&0\\ F^T KF&0&-I/\rho \end{array}\right]<0
\end{equation}
where $B=A^T X+XA+2F^TY+2Y^TF+E^T E/\gamma^2+R$ and $A=-i\Sigma M_P-\frac{1}{2}\Sigma N_P^\dag JN_P$, then the associated cost function satisfies the bound
\begin{equation}\label{optimal_controller}
\begin{split}
&\limsup \limits_{T\to \infty} \frac{1}{T}\int_0^T\langle W(t)\rangle dt\leq\lambda+\delta,
\end{split}
\end{equation}
where $W(t)$ is in the form of (\ref{nondynamic_wt}) and
\begin{equation}
\lambda=\text{Tr}(X\Sigma N_P^\dag \left[\begin{array}{cc} I&0\\0&0\end{array}\right]N_P\Sigma).
\end{equation}
\end{theorem}
\emph{Proof:}
Suppose the inequality (\ref{nondynamic_inequality}) is satisfied. Using the Schur complement \cite{Boyd}, we have \begin{equation}\label{nondynamic_inequality1}
\left[\begin{array}{ccc}B+16X\theta^TE^TE\theta X&F^T KF\\ F^T KF&-I/\rho \end{array}\right]<0.
\end{equation}
Also, it follows from the Schur complement that (\ref{nondynamic_inequality1}) is equivalent to the following inequality
\begin{equation}\label{nondynamic_inequality2}
\begin{split}
&A^T X+XA+2F^TY+2Y^TF+E^T E/\gamma^2\\
&+16X\theta^TE^TE\theta X
+R+\rho F^T KFF^T KF<0.
\end{split}
\end{equation}
Substituting $Y=KF\theta^T X$ into the inequality (\ref{nondynamic_inequality2}), we obtain that
\begin{equation}\label{0401}
\begin{split}
&(A-i\Sigma F^T K F)^T X+X(A-i\Sigma F^TKF)\\
&+4X\Sigma^\dag E^TE\Sigma X
+E^T E/\gamma^2+R+\rho F^T KFF^T KF<0.
\end{split}
\end{equation}
It follows straightforwardly from (\ref{0401}) that $A-i\Sigma F^T K F$ is Hurwitz.
Based on Lemma \ref{algebra} and Lemma \ref{zlemma}, we have
\begin{equation}
\begin{split}
&-i[V,H_P+H_{K1}]+\mathcal{L}(V)+[V,z^T][z,V]+\frac{1}{\gamma^2}z^T z+W\\
&=\left[\begin{array}{l}q\\p\end{array}\right]^T  \left(\begin{array}{l}(A-i\Sigma F^TK F)^T X\\
+X(A-i\Sigma F^TKF)\\
+4X\Sigma^\dag E^T E\Sigma X+E^T E/\gamma^2\\
+R
+\rho F^T KFF^T KF\end{array}\right) \left[\begin{array}{l}q\\p\end{array}\right]\\
&\quad +\text{Tr}(X\Sigma N_P^\dag \left[\begin{array}{cc} I&0\\0&0\end{array}\right]N_P\Sigma).
\end{split}
\end{equation}
It follows from Lemma \ref{lemmaw1} and Lemma \ref{lemma_perturbation} that the following conditions are satisfied;
\begin{equation}
\limsup \limits_{T\to \infty} \frac{1}{T}\int_0^T\langle W(t)\rangle dt\leq \lambda+\delta,
\end{equation}
where
$\lambda= \text{Tr}(X\Sigma N_P^\dag \left[\begin{array}{cc} I&0\\0&0\end{array}\right]N_P\Sigma)$.
\hfill $\Box$
\begin{remark}
In order to construct the guaranteed cost dynamic controller in Theorem \ref{nondynamic_theorem}, we should not only satisfy the linear matrix inequality (\ref{nondynamic_inequality}), but also guarantee the equality $Y=KF\theta^T X$. We should notice that the above equality is non-convex constraint. Hence, we will convert this problem into a rank constrained LMI problem. The numerical method is similar as that in section \ref{Dynamic Coherent Controller}-B. Thus a detailed procedure is omitted in this section.
\end{remark}
\section{Dynamic Coherent Controller}\label{Dynamic Coherent Controller}
In this section, we design a dynamic coherent controller that is realized by being directly coupled to the original uncertain quantum system. The directly coupled controller and the quantum plant may interact with each other by exchanging energy via an interaction Hamiltonian.
We should also note that the associated operators of the nominal system commute with all the associated operators of the dynamic controller. In the following part, we will introduce the design of dynamic quantum controllers and then formulate the augmented quantum system consisting of the quantum plant and the directly coupled quantum controller using an $(S,L,H)$ parameterization.

We consider a linear dynamic quantum controller Hamiltonian in the form of
\begin{equation}\label{H_4}
H_{K2}=\frac{1}{2}\left[\begin{array}{c} q_K \\p_K\end{array}\right]^T K_{22}\left[\begin{array}{l} q_K \\ p_K\end{array}\right],
\end{equation}
where $K_{22}\in \mathbb{R}^{ 2n_K\times 2n_K}$.
In what follows, we take the interaction Hamiltonian to be
\begin{equation}\label{H_5}
\begin{split}
H_{int}=&\frac{1}{2}\left[\begin{array}{c} q_K \\p_K\end{array}\right]^T  K_{12}^T F\left[\begin{array}{l} q \\ p\end{array}\right]\\
&+\frac{1}{2}\left[\begin{array}{c} q \\p\end{array}\right]^T F^T K_{12} \left[\begin{array}{l} q_K \\ p_K\end{array}\right],\\
\end{split}
\end{equation}
where $K_{12}\in \mathbb{R}^{ 2n_P\times 2n_K}$. Also, we consider a static coherent controller in the form of (\ref{H3}).

It follows from (\ref{H3}), (\ref{H_4}) and (\ref{H_5}) that the total controller Hamiltonian is in the form of
\begin{equation}\label{total__controller_Hamiltonian}
\begin{split}
H_K&=H_{int}+H_{K1}+H_{K2}\\
&=\frac{1}{2}\left[\begin{array}{l}q\\p\\q_K\\p_K\end{array}\right]^T\left[\begin{array}{ll}F^T K_{11} F&F^T K_{12}\\K_{12}^TF&K_{22}\end{array}\right]\left[\begin{array}{l}q\\p\\q_K\\p_K\end{array}\right].
\end{split}
\end{equation}
Here, we denote $K=\left[\begin{array}{ll}K_{11}&K_{12}\\K_{12}^T&K_{22}\end{array}\right]$,
and
$\overline{F}=\left[\begin{array}{ll}F&0\\0&I\end{array}\right]$. Consequently,
the total Hamiltonian for the augmented quantum linear system (excluding the perturbation Hamiltonian) that comprises the nominal quantum plant and the quantum controller is
\begin{equation}\label{total_Hamiltonian}
H_{P+K}=H_P+H_K.
\end{equation}

We know that the coupling operator $L_P$ and $\left[\begin{array}{c} L_P \\ L_P^\#\end{array}\right]$ for the plant are defined in (\ref{single_coupling}) and (\ref{coupling}), respectively.
The coupling operator for the controller is assumed to be in the form
\begin{equation}\label{dynamic_couple}
L_K=\left[\begin{array}{cc} N_{K1} & N_{K2}\end{array}\right]\left[\begin{array}{l} q_K \\ p_K\end{array}\right],
\end{equation}
where $N_{K1}\in\mathbb{C}^{m_K\times n_K}$ and $N_{K2}\in\mathbb{C}^{m_K\times n_K}$ .\\
Also, we have
\begin{equation}
\left[\begin{array}{c} L_K \\ L_K^\#\end{array}\right]=N_K\left[\begin{array}{l} q_K \\ p_K\end{array}\right]=\left[\begin{array}{cc} N_{K1} & N_{K2}\\N_{K1}^\# & N_{K2}^\#\end{array}\right]\left[\begin{array}{l} q_K \\ p_K\end{array}\right].
\end{equation}
Hence, the overall coupling operator for the directly coupled system comprising the plant and the controller is of the form
\begin{equation}\label{total_coupling}
L=\left[\begin{array}{c} L_P \\ L_K\end{array}\right]=\left[\begin{array}{cccc} N_{P1} & N_{P2}&0&0\\0&0&N_{K1} & N_{K2}\end{array}\right]\left[\begin{array}{l}  q \\ p\\q_K \\ p_K\end{array}\right].
\end{equation}
Also, we write
\begin{equation}\label{total_L}
\begin{split}
\left[\begin{array}{c} L \\ L^\#\end{array}\right]=&\left[\begin{array}{c} L_P \\ L_K\\L_P^\# \\ L_K^\#\end{array}\right]=N\left[\begin{array}{l}  q \\ p\\q_K \\ p_K\end{array}\right]\\
=&\left[\begin{array}{cccc} N_{P1} & N_{P2}&0&0\\0&0&N_{K1} & N_{K2}\\N_{P1}^\# & N_{P2}^\#&0&0\\0&0&N_{K1}^\# & N_{K2}^\#\end{array}\right]\left[\begin{array}{l}  q \\ p\\q_K \\ p_K\end{array}\right].
\end{split}
\end{equation}

The ``Lyapunov" operator $V$ for the augmented system is considered to be a non-negative self-adjoint operator of the form
\begin{equation}\label{v_dynamic}
V=\left[\begin{array}{l}q\\p\\q_K\\p_K\end{array}\right]^T \left[\begin{array}{cc}X_{11}&X_{12}\\X_{12}^T&X_{22}\end{array}\right]\left[\begin{array}{l}q\\p\\q_K\\p_K\end{array}\right]
\end{equation}
where $X=\left[\begin{array}{cc}X_{11}&X_{12}\\X_{12}^T&X_{22}\end{array}\right]\in \mathbb{R}^{2(n_P+n_K)\times2(n_P+n_K)}$  is a symmetric positive definite matrix.

Hence, a set of non-negative self-adjoint operators $\mathcal{P}$ can be defined as
\begin{equation}
\tilde{\mathcal{P}}=\left\{\begin{array}{c} V\ \textnormal{ of the form} \ (\ref{v_dynamic})\ \textnormal{where} \ X>0\ \textnormal {is a}\\ \textnormal{symmetric positive definite matrix}\end{array} \right\}.
\end{equation}
Since all variables of the plant are assumed to commute with all variables of the controller, we have the following commutation relation
\begin{equation}\label{commutation1}
\begin{split}
&\left[\begin{array}{c}\left[\begin{array}{l}q\\p\\q_K\\p_K\end{array}\right] ,\left[\begin{array}{l}q\\p\\q_K\\p_K\end{array}\right]^T \end{array}\right]
=2i\Theta=\Xi
\end{split}
\end{equation}
where $\Theta=\left[\begin{array}{cr}\theta&0\\0&\theta \end{array} \right]$ .
In order to proceed to the following section, we introduce the permutation matrix $P_{n+m}$, where the symbol $P_{n+m}$ refers to a $2(n+m)\times2(n+m)$ matrix. The permutation matrix $P_{n+m}$ is defined in such way that if we consider a column vector $a=[a_1\quad a_2\quad ...\quad a_{n+m}\quad...\quad a_{2(n+m)}]^T$, then $P_{n+m}a=[a_1\quad a_2\quad ...\quad a_n\quad a_{n+m+1}\quad a_{n+m+2}\quad ...\quad\\
 a_{2n+m}\quad a_{n+1}\quad a_{n+2}\quad ...\quad a_{n+m}\quad a_{2n+m+1}\quad ...\\
 \quad a_{2(n+m)}]^T.$
 Recall the property of an $m\times m$ permutation matrix that it is a full-rank real matrix whose columns comprise standard basis vector for $\mathbb{R}^m$, that is, vectors in $\mathbb{R}^m$ contains precisely a single element with value 1 and all the remaining elements are 0. A permutation matrix $P$ also
 has the unitary property $PP^T=P^TP=I$. Another notation we need to introduce is $J_n=\left[\begin{array}{cc}I_n&0\\0&-I_n\end{array}\right]$.
Hence, we have
\begin{equation}\label{permutation}
\begin{split}
&P_{m_P+m_K}N=\left[\begin{array}{ll}N_P&0\\0&N_K\end{array}\right],\\
&N=P_{m_P+m_K}^T\left[\begin{array}{ll}N_P&0\\0&N_K\end{array}\right],\\
&P_{m_P+m_K}^T J_{m_P+m_K} P_{m_P+m_K}=\left[\begin{array}{cc}J_{m_P}&0\\0&J_{m_K}\end{array}\right].
\end{split}
\end{equation}


In order to present the main result, we require some algebraic identities.

\begin{lemma}\label{algebra_dynamic}
Suppose $\mathcal{V}\in\tilde{\mathcal{P}}$, $H_{P+K}$ is in the form of (\ref{total_Hamiltonian}) and $L$ is in the form of (\ref{total_L}). Then, we have that
\begin{equation}
\begin{split}
-&i[V,H_{P+K}]+\mathcal{L}(V)\\
=&\left[\begin{array}{l}q\\p\\q_K\\p_K\end{array}\right]^T  \left(\begin{array}{l}(\overline{A}-i\Xi \overline{F}^T K\overline{F})^T X\\
+X(\overline{A}-i\Xi \overline{F}^T K\overline{F})
\end{array}\right) \left[\begin{array}{l}q\\p\\q_K\\p_K\end{array}\right]\\
&+\textnormal{Tr}(X\Xi N^\dag \left[\begin{array}{cc} I&0\\0&0\end{array}\right]N\Xi),
\end{split}
\end{equation}
\begin{equation}\label{overline_A}
\textnormal{where}\quad \overline{A}=\left[\begin{array}{cc}-i\Sigma M_P-\frac{1}{2}\Sigma N_P^\dag JN_P&0\\0&-\frac{1}{2}\Sigma N_K^\dag JN_K\end{array}\right].
\end{equation}
\end{lemma}
\emph{Proof:} According to Lemma \ref{algebra}, we know that
\begin{equation}\label{A1}
\begin{split}
-&i[V,H_{P+K}]+\mathcal{L}(V)\\
=&\left[\begin{array}{l}q\\p\\q_K\\p_K\end{array}\right]^T  \left(\begin{array}{l}(-i\Xi M-\frac{1}{2}\Xi N^\dag JN)^T X\\
+X(-i\Xi M-\frac{1}{2}\Xi N^\dag JN)
\end{array}\right) \left[\begin{array}{l}q\\p\\q_K\\p_K\end{array}\right]\\
&+\textnormal{Tr}(X\Xi N^\dag \left[\begin{array}{cc} I&0\\0&0\end{array}\right]N\Xi),
\end{split}
\end{equation}
where $M=\left[\begin{array}{ll}M_P+F^T K_{11} F&F^T K_{12}\\K_{12}^TF&K_{22}\end{array}\right]$.
To shorten the writing, we denote $\bold{A}=-i\Xi M-\frac{1}{2}\Xi N^\dag JN$ and we have

\begin{equation}
\begin{split}
\bold{A}=&-i\Xi\left[\begin{array}{ll}M_P+F^T K_{11} F&F^T K_{12}\\K_{12}^TF&K_{22}\end{array}\right]\\
&-\frac{1}{2}\Xi N^\dag
J_{m_P+m_K}N\\
\end{split}
\end{equation}
Following (\ref{permutation}), we get
\begin{equation}\label{A}
\begin{split}
\bold{A}=&-i\Xi\left[\begin{array}{ll}M_P+F^T K_{11} F&F^T K_{12}\\K_{12}^TF&K_{22}\end{array}\right]\\
&-\frac{1}{2}\Xi\left[\begin{array}{ll}N_P^\dag&0\\0&N_K^\dag\end{array}\right]
\left[\begin{array}{cc}J_{m_P}&0\\0&J_{m_K}\end{array}\right]\left[\begin{array}{ll}N_P&0\\0&N_K\end{array}\right]\\
=&\left[\begin{array}{cc}-i\Sigma M_P-\frac{1}{2}\Sigma N_P^\dag JN_P&0\\0&-\frac{1}{2}\Sigma N_K^\dag JN_K\end{array}\right]\\
&-i\Xi \overline{F}^T K\overline{F}\\
=&\overline{A}-i\Xi \overline{F}^T K\overline{F}.
\end{split}
\end{equation}
Therefore, the result follows from (\ref{A1}) and (\ref{A}).
\hfill $\Box$

We know that only the nominal quantum system is subject to uncertain quantum perturbation. Hence, we rewrite the quadratic perturbation uncertainty (\ref{perturbation_equation}) and (\ref{z}) in the following form
\begin{equation}\label{H2_dynamic}
H_{un}=\frac{1}{2}\left[\begin{array}{l}q\\p\\q_K\\p_K\end{array}\right]^T \left[\begin{array}{ll}E&0\end{array}\right]^T\Delta \left[\begin{array}{ll}E&0\end{array}\right]\left[\begin{array}{l}q\\p\\q_K\\p_K\end{array}\right],
\end{equation}
\begin{equation}\label{z_dynamic1}
z=\left[\begin{array}{l} \zeta_q \\ \zeta_p\end{array}\right]=\left[\begin{array}{ll} E & 0\end{array}\right] \left[\begin{array}{llll} q^T & p^T&q_K^T&p_K^T\end{array}\right]^T.
\end{equation}
In order to better present the following theorem, we denote $\overline{E}=\left[\begin{array}{ll}E&0\end{array}\right]$.


Now we are in the position to define an associated cost function as
\begin{equation}
J=\limsup \limits_{T\to \infty} \frac{1}{T}\int_0^\infty \langle W\rangle dt, \quad \text{and}
\end{equation}
\begin{equation}\label{dynamic_wt}
\begin{split}
W=&\left[\begin{array}{l}q\\p\end{array}\right]^T R \left[\begin{array}{l}q\\p\end{array}\right]+\left[\begin{array}{l}q\\p\\q_K\\p_K\end{array}\right]^T\rho\overline{F}^T K\overline{F}\overline{F}^T K\overline{F}\left[\begin{array}{l}q\\p\\q_K\\p_K\end{array}\right]\\
=&\left[\begin{array}{l}q\\p\\q_K\\p_K\end{array}\right]^T(\overline{R}+\rho\overline{F}^T K\overline{F}\overline{F}^T K\overline{F})\left[\begin{array}{l}q\\p\\q_K\\p_K\end{array}\right],
\end{split}
\end{equation}
where $\overline{R}=\left[\begin{array}{ll}R&0\\0&0\end{array}\right]$ and $\rho\in(0,\infty)$ is a weighting factor.
\subsection{Controller Design}
In this subsection, we present the main result on dynamic coherent controller design.
\begin{theorem}\label{dynamic_controller_design}
Consider an uncertain quantum system $(S_P, L_P, H_{P+un})$, where $H_{P+un}=H_P+H_{un}$, $H_P$ is in the form of (\ref{Hamiltonian}),
$L_P$ is of the form (\ref{single_coupling}) and $H_{un}\in\mathcal{W}_2$. It is directly coupled with a dynamic quantum controller $(S_K, L_K, H_{K})$.
 Then the closed loop quantum system is defined by $(S,L,H)$ where $H=H_P+H_{un}+H_K$, $H_K$ is in the form of (\ref{total__controller_Hamiltonian}) and $L$ is in the form of (\ref{total_coupling}). If there exists a symmetric matrix $K$, $X>0$ and $Y=K\overline{F}\Theta^T X$ such that
\begin{equation}\label{dynamic_inequality}
\left[\begin{array}{ccc}B&4X\Theta^T\overline{E}^T&\overline{F}^TK\overline{F}\\4\overline{E}\Theta X&-I&0\\ \overline{F}^T K\overline{F}&0&-I/\rho \end{array}\right]<0
\end{equation}
where $B=\overline{A}^T X+X\overline{A}+2\overline{F}^TY+2Y^T\overline{F}+\overline{E}^T \overline{E}/\gamma^2+\overline{R}$ and $\overline{A}$ is in the form of (\ref{overline_A}), then the associated cost function satisfies the bound
\begin{equation}\label{optimal_controller}
\begin{split}
&\limsup \limits_{T\to \infty} \frac{1}{T}\int_0^T\langle W(t)\rangle dt\leq\lambda+\delta,
\end{split}
\end{equation}
where $W(t)$ is in the form of (\ref{dynamic_wt}) and
\begin{equation}\label{cost_bound}
\lambda=\text{Tr}(X\Xi N^\dag \left[\begin{array}{cc} I&0\\0&0\end{array}\right]N\Xi).
\end{equation}
\end{theorem}
\emph{Proof:}
Suppose the conditions of the theorem are satisfied. Using the Schur complement, (\ref{dynamic_inequality}) is equivalent to
\begin{equation}\label{dynamic_inequality1}
\left[\begin{array}{ccc}B+16X\Theta^T\overline{E}^T\overline{E}\Theta X&\overline{F}^TK\overline{F}\\ \overline{F}^T K\overline{F}&-I/\rho \end{array}\right]<0.
\end{equation}
Applying the Schur complement to (\ref{dynamic_inequality1}), we obtain
\begin{equation}\label{dynamic_inequality2}
\begin{split}
&\overline{A}^T X+X\overline{A}+2\overline{F}^TY+2Y^T\overline{F}+\overline{E}^T \overline{E}/\gamma^2\\
&+16X\Theta^T\overline{E}^T\overline{E}\Theta X
+\overline{R}+\rho\overline{F}^TK\overline{F}\overline{F}^T K\overline{F}<0.
\end{split}
\end{equation}
Substituting $Y=K\overline{F}\Theta^T X$  into (\ref{dynamic_inequality2}), we have
\begin{equation}\label{04011}
\begin{split}
&(\overline{A}-i\Xi\overline{F}^TK\overline{F})^T X+X(\overline{A}-i\Xi\overline{F}^TK\overline{F})\\
&+4X\Xi^\dag\overline{E}^T\overline{E}\Xi X
+\overline{E}^T \overline{E}/\gamma^2+\overline{R}+\rho\overline{F}^TK\overline{F} \overline{F}^T K\overline{F}<0
\end{split}
\end{equation}
It follows straightforwardly from (\ref{04011}) that $\overline{A}-i\Xi\overline{F}^TK\overline{F}$ is Hurwitz.
Following Lemma \ref{zlemma}, (\ref{H2_dynamic}) and (\ref{z_dynamic1}), we have
\begin{equation}\label{Vz}
\begin{split}
&[V,z^T][z,V]+\frac{1}{\gamma^2}z^T z\\
&=\left[\begin{array}{l}q\\p\\q_K\\p_K\end{array}\right]^T\left(\begin{array}{l} 4X\Xi^\dag \overline{E}^T \overline{E}\Xi X\\+\overline{E}^T \overline{E}/\gamma^2\end{array}\right)\left[\begin{array}{l}q\\p\\q_K\\p_K\end{array}\right].
\end{split}
\end{equation}
Based on (\ref{Vz}) and Lemma \ref{algebra_dynamic}, the following equation is achieved;
\begin{equation}
\begin{split}
&-i[V,H_{P+K}]+\mathcal{L}(V)+[V,z^T][z,V]+\frac{1}{\gamma^2}z^T z+W\\
&=\left[\begin{array}{l}q\\p\\q_K\\p_K\end{array}\right]^T  \left(\begin{array}{l}(\overline{A}-i\Xi\overline{F}^TK\overline{F})^T X\\
+X(\overline{A}-i\Xi\overline{F}^TK\overline{F})\\
+4X\Xi^\dag\overline{E}^T\overline{E}\Xi X\\
+\overline{E}^T \overline{E}/\gamma^2+\overline{R}\\
+\rho\overline{F}^TK\overline{F} \overline{F}^T K\overline{F}\end{array}\right) \left[\begin{array}{l}q\\p\\q_K\\p_K\end{array}\right]\\
&\quad +\text{Tr}(X\Xi N^\dag \left[\begin{array}{cc} I&0\\0&0\end{array}\right]N\Xi).
\end{split}
\end{equation}
According to Lemma \ref{lemma_perturbation} and Lemma \ref{lemmaw1}, we obtain
\begin{equation}
\limsup \limits_{T\to \infty} \frac{1}{T}\int_0^T\langle W(t)\rangle dt\leq \lambda+\delta,
\end{equation}
where
$\lambda= \text{Tr}(X\Xi N^\dag \left[\begin{array}{cc} I&0\\0&0\end{array}\right]N\Xi)$.

\hfill $\Box$

\subsection{Numerical Procedure}
Theorem \ref{dynamic_controller_design} can be formulated as the minimization of the cost bound $\lambda$ subject to the LMI (\ref{dynamic_inequality}) and the equality $Y=K\overline{F}\Theta^T X$ constraint. Due to a fact that the equality constraint is not linear, we cannot use a traditional LMI technique. Therefore, this problem is transformed into a rank constrained LMI problem \cite{Orsi2}. The way to deal with this problem is to linearize the equality $Y=K\overline{F}\Theta^T X$ by introducing appropriate matrix lifting variables and the associated equality constraints, and transforming them into an LMI with a rank constraint.

We set $Z_{x1}=X, Z_{x2}=K$ and introduce appropriate matrix lifting variables $Z_{v1}=K\overline{F}\Theta^T , Z_{v2}=K\overline{F}\Theta^T X$. Then we define $Z$ as a symmetric matrix with dimension $5(2n_P+2n_K)\times 5(2n_P+2n_K)$ and $Z=VV^T$, where $V=[I\quad Z_{x1}^T\quad Z_{x2}^T\quad Z_{v1}^T\quad Z_{v2}^T]^T$.

We require the symmetric matrix $Z$ to satisfy the following conditions:
\begin{equation}\label{nondynamic_equality}
\begin{array}{ll}
Z\geq0;                           &Z_{0,0}-I=0;\\
Z_{x1}-Z_{x1}^T=0;                &Z_{x2}-Z_{x2}^T=0;\\
Z_{v1}-Z_{x2}\overline{F}\Theta^T=0; &Z_{v2}-Z_{v1}Z_{x1}=0
\end{array}
\end{equation}
and a rank constraint
\begin{equation}\label{nondynamic_rank}
\text{rank}(Z)\leq 2n_P+2n_K.
\end{equation}

Therefore, the way to realize Theorem \ref{dynamic_controller_design} is to solve (\ref{dynamic_inequality})
subject to condition (\ref{nondynamic_equality}) and (\ref{nondynamic_rank}) using a rank constrained LMI.
To solve the rank constrained LMI problem above, we work in Matlab using an algorithm \cite{Orsi1} and also use the Yalmip optimization prototyping environment \cite{Yalmip} and SeDuMi solver \cite{sedumi}.

In order to minimize the cost bound $\lambda$, we also need to introduce another inequality
\begin{equation}
\text{Tr}(X\Xi N^\dag \left[\begin{array}{cc} I&0\\0&0\end{array}\right]N\Xi)<\beta.
\end{equation}
Hence, to minimize the cost bound, we just need to lower the pre-specified bound $\beta$ using the bisection method; e.g., \cite{bisection}.
\section{Illustrative Example}\label{Illustrative Example}

In order to illustrate the controller design methods demonstrated in this paper, we are now present an example. We consider an uncertain quantum system defined by a triple $(S,L,H)$ in the following form
\begin{equation}
\begin{split}
&S_P=I,L_P=\frac{1}{2}\left[\begin{array}{cr} \sqrt{\kappa}&\sqrt{\kappa}i\\ \end{array}\right]\left[\begin{array}{l} q \\ p\end{array}\right],\\
&H_{P+{un}}=\frac{1}{2}\left[\begin{array}{c c} q^T & p^T\end{array}\right]\left[\begin{array}{cc} 0.7&1 \\ 1&1.8\end{array}\right]\left[\begin{array}{l} q \\ p\end{array}\right],\nonumber
\end{split}
\end{equation}
where $H_{P+{un}}=H_P+H_{un}$. The Hamiltonian of the nominal system is
\begin{equation}
H_P=\frac{1}{2}\left[\begin{array}{c c} q^T & p^T\end{array}\right]\left[\begin{array}{cc} -0.55&0 \\ 0&0.55\end{array}\right]\left[\begin{array}{l} q \\ p\end{array}\right],\nonumber
\end{equation}
and the Hamiltonian of the perturbation system is
\begin{equation}
H_{un}=\frac{1}{2}\left[\begin{array}{c c} q^T & p^T\end{array}\right]\left[\begin{array}{cc} 1.25&1 \\ 1&1.25\end{array}\right]\left[\begin{array}{l} q \\ p\end{array}\right].\nonumber
\end{equation}
This leads to the corresponding parameters in Theorem \ref{nondynamic_theorem} and Theorem \ref{dynamic_controller_design} as follows:
\begin{equation}
\begin{split}
&M_P=\left[\begin{array}{cc} -0.55&0 \\ 0&0.55\end{array}\right],
N_P=\frac{1}{2}\left[\begin{array}{cc} \sqrt{\kappa}&\sqrt{\kappa}i\\
\sqrt{\kappa}&-\sqrt{\kappa}i\end{array}\right],\\
&N_K=\frac{1}{2}\left[\begin{array}{cc} \sqrt{0.5}&\sqrt{0.5}i\\
\sqrt{0.5}&-\sqrt{0.5}i\end{array}\right], F=\left[\begin{array}{cc} 1&0\\ 0&1\end{array}\right],\\
&A=\left[\begin{array}{cc} -\frac{\kappa}{2}&1.1\\ 1.1&-\frac{\kappa}{2}\end{array}\right],
E=\left[\begin{array}{cc} 1&0.5\\ 0.5&1\end{array}\right],\Delta=\left[\begin{array}{cc} 1&0\\ 0&1\end{array}\right].\nonumber
\end{split}
\end{equation}

By running the program, we find that the nominal plant is mean square stable for the range $\kappa\in(8.2,\infty)$ without any controller.
Nevertheless, a static coherent controller can stabilize the system for the range of $\kappa\in(7.1,\infty)$.
Meanwhile, a dynamic coherent controller can guarantee that the system is stable for the range of $\kappa\in(7,\infty).$ Therefore, we can conclude that quantum controllers can stabilize uncertain quantum systems for a larger range of $\kappa$ than the one without a controller.

With regard to the stability and the performance of the quantum systems, the importance of the static controller to the given system has already been addressed in \cite{Chengdi1}. Here, we will focus our attention on a performance comparison between the given system with a static quantum controller and that with a dynamic quantum controller. As displayed in Figure 1, compared to the static controller, the dynamic quantum controller can guarantee the uncertain quantum system stable for a larger range of $\kappa$ values with a lower cost. Hence, it leads to a closed loop system having better performance.
\begin{figure}[htb]
       \centering
        \includegraphics[width=0.5  \textwidth]{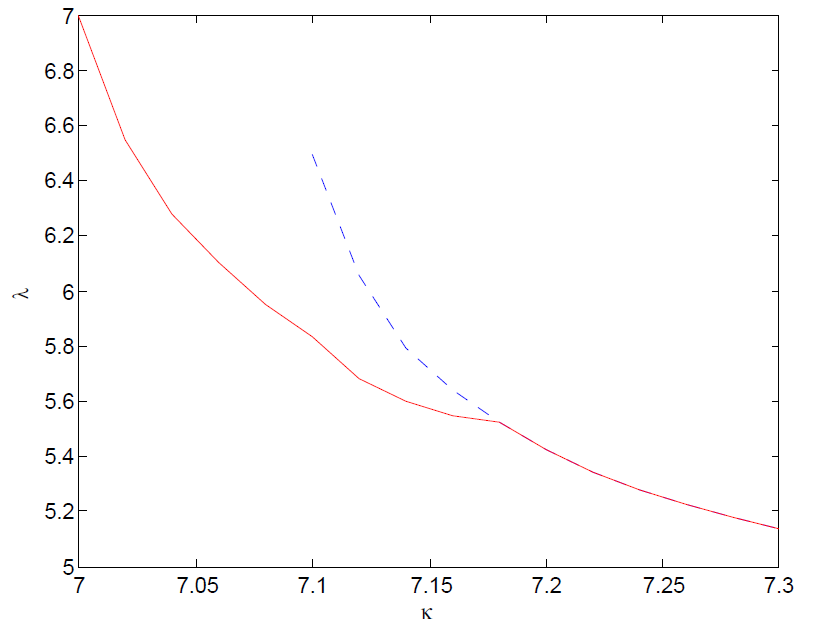}
        \caption{Guaranteed cost bounds for uncertain quantum systems with a dynamic controller (solid line) and with a static controller (dash line)}
        \label{fig1}
\end{figure}
\section{Conclusions}\label{conclusion}

This paper has evaluated the performance of uncertain linear quantum systems subject to quadratic perturbations in the system Hamiltonian. We proposed methods for a static coherent quantum controller design and a dynamic coherent quantum controller design. We converted the controller design problem into a rank constrained LMI problem and solved the problem based on an alternating projections algorithm. An example was presented to illustrate these two controller design methods. By comparison, we showed that a dynamic controller can have an improved control performance over a static one.


\end{document}